\documentclass[aps,reprint,superscriptaddress,floatfix,prl]{revtex4-1}

\usepackage{graphicx}
    \graphicspath{{./Images/}}
\usepackage{xcolor}
\usepackage[colorlinks = true, citecolor=blue, linkcolor=black]{hyperref}
\usepackage{enumitem}
    \setlist[itemize]{noitemsep, topsep=0pt}
    \setlist[enumerate]{noitemsep, topsep=0pt}
\usepackage{verbatim}
\usepackage{mathtools,amssymb,amsmath}
\usepackage{bm}
\usepackage{yhmath}
\usepackage[inline]{asymptote}
\usepackage{cancel}
\usepackage{relsize}
\usepackage{array}
\usepackage{bm}

\newcommand{\be}{\begin{equation}}
\newcommand{\ee}{\end{equation}}
\newcommand{\<}{\langle}
\renewcommand{\>}{\rangle}
\newcommand{\ve}{\varepsilon}
\newcommand{\Tr}{{\rm Tr\,}}
\newcommand{\tr}{{\rm tr\,}}
\usepackage{color}

\renewcommand{\vec}[1]{{\bf #1}}
\DeclareMathOperator*{\Motimes}{\text{\raisebox{0.25ex}{\scalebox{0.8}{$\bigotimes$}}}}

\begin{document}

\title{Unitary Subharmonic Response and Floquet Majorana Modes}

\author{Oles Shtanko}
\affiliation{Joint Quantum Institute, NIST/University of Maryland, College Park, MD 20742, USA}
\affiliation{Joint Center for Quantum Information and Computer Science, NIST/University of Maryland, College Park, MD 20742, USA}

\author{Ramis Movassagh}
\affiliation{IBM Research,  MIT-IBM AI lab,  Cambridge MA, 02142, USA}

\begin{abstract}
Detection and manipulation of excitations with non-Abelian statistics, such as Majorana fermions, are essential for creating topological quantum computers. To this end, we show the connection between the existence of such localized particles and the phenomenon of unitary subharmonic response (SR) in periodically driven systems. In particular, starting from highly nonequilibrium initial states, the unpaired Majorana modes exhibit spin oscillations with twice the driving period, are localized, and can have exponentially long lifetimes in clean systems. While the lifetime of SR is limited in translationally invariant systems, we show that disorder can be engineered to stabilize the subharmonic response of Majorana modes. A viable observation of this phenomenon can be achieved using modern multiqubit hardware, such as superconducting circuits and cold atomic systems.
\end{abstract}

\maketitle

 The recent experimental frontiers have succeeded in the creation and manipulation of systems consisting of many well-isolated controllable qubits \cite{bernien2017probing,zhang2017observation,zeiher2016many,endres2016atom}. Such devices promise to have applications from quantum computing to simulating quantum many-body systems out of equilibrium \cite{georgescu2014quantum,chang2018quantum,polkovnikov2011nonequilibrium,gross2017quantum}. In the context of periodically driven systems, the prominent examples of such nonequilibrium systems are ones exhibiting persistent oscillations with a period equal to multiple initial driving periods. This phenomenon was recently studied in the context of discrete time crystals \cite{else2016floquet, else2019discrete,else2017prethermal,yao2017discrete,sacha2017time,moessner2017equilibration,khemani2016phase, von2016absolute,ho2017critical}
and reported in several experimental settings \cite{zhang2017j,choi2017observation,rovny2018observation,pal2018temporal}.

In this Letter, we study the oscillations similar to time crystals but localized only at the boundaries of symmetry-protected topological (SPT) phases \cite{kitagawa2010topological,vonkeyserlingk2016phase,roy2017periodic}. In equilibrium, one-dimensional SPT phases are widely studied due to the emergence of topologically protected Majorana zero modes (MZM) at the boundaries \cite{kitaev2001unpaired,bahri2015topological}. This phenomenon is of interest for fundamental physics perspective and the potential for realization of robust quantum computing \cite{nonabelian2008nayak}. In a driven setting, SPT phases have even richer phenomenology and may exhibit, additionally to MZM, a pair of Majorana $\pi$ modes (MPM) \cite{jiang2011majorana,liu2013floquet,potter2016classification,floquet2017potirniche}. Here we study in detail how the emergence of MPM, in connection with MZM, leads to robust  double-period oscillations at the boundaries \cite{bomantara2018simulation},  that can be generalized to models with triple and larger periodicities \cite{sreejith2016parafermion,chew2020time}. We also show that multiperiod boundary oscillations may exhibit a sufficiently long equilibration time resisting thermalization  \cite{else2017prethermal,weidinger2017floquet,abanin2017effective,abanin2017rigorous,else2017prethermal2} and can be reliably protected by a mechanism of many-body localization (MBL) \cite{ponte2015many,zhang2016floquet,ponte2015periodically}. We also propose the boundary double-period oscillations as an alternative probe of Floquet Majorana modes in quantum systems. In particular, the observation of local persistent two-period oscillations can be used to establish both the presence of Majorana modes, their physical location, and localization length.  

We consider the Majorana modes oscillations in a broader context of subharmonic response (SR) defined in the following way. 
Consider a Floquet system defined by a time-dependent Hamiltonian $H(t) = H(t+T)$, where $T$ is a fixed period. The periodic field affects the time dependence of
expectations for local observables operators $O_\mu$(t).
We define SR as a phenomenon when one or several of these observables permanently oscillate with a period $kT$ for integer $k>1$, i.e., $\<O_\mu(t)\>=\<O_\mu(t+kT)\>$, where $\<\dots\>$ is the expectation value in the initial state. At the same time, the SR oscillations must persist regardless of the choice of the initial state and in the presence of small but finite perturbations.
Therefore, by definition, SR does not include fine-tuned systems, for example, synchronized uncoupled qubits and integrable systems, due to lack of robustness to factors such as disorder, qubit coupling, or initial conditions. Also, SR is more broadly defined than the discrete time crystal because it does not require long-range spatial correlations across the system. 

Under what conditions does SR happen? To address this question, it is convenient to limit our consideration from continuous time $t$ to the discrete stroboscopic time $t_n = nT$. The discrete time dynamics is generated by the unitary Floquet operator $
U_F = \mathcal T \exp \Bigl(-i\int_0^TH(t)\:dt\Bigl)$
describing time evolution between discrete times $t_n$ and $t_{n+1}$.  Then, a sufficient condition for SR of observable $O_\mu$ is the existence of a set of local oscillating conserved unitary modes $\tau_i$ that exhibit $U_F\tau_i U_F^\dag = e^{2\pi i/k}\tau_i$ and can be measured via observation of $O_\mu$, i.e., $|{\rm tr}\, O_\mu \tau_i|>0$, where ${\rm tr(\dots)}$ is normalized trace (see Supplemental Material for discussion ).
The robust appearance of conserved oscillating modes requires specific symmetries.  

Floquet SPT phases are examples for which symmetries lead to the emergence of localized oscillating modes at system's boundaries. For example, as an illustration we study a toy SPT model 
 of periodically driven Ising-type system exhibiting Majorana modes. This system can be mapped to free fermions with Floquet operator $U_F$ that satisfies $U_F\Gamma^\alpha_i  = e^{i\alpha}\Gamma^\alpha_iU_F$, where $\alpha$ = $0$, $\pi$, operator $\Gamma^0_i$ is MZM, and operator $\Gamma^\pi_i$ is MPM. Majorana modes satisfy $\Gamma^\alpha_i\Gamma^\beta_j+\Gamma^\beta_j\Gamma^\alpha_i = 2\delta_{ij}\delta_{\alpha\beta}$, and localized at the boundaries of the 1D system. This is enforced if the Floquet operator of the system commutes with the parity operator $\mathcal P$ 
 \cite{potter2016classification}. 
 The oscillating modes can be constructed as $\Gamma^\pi_i$, $\mathcal P\Gamma^\pi_i$, and $\Gamma^\pi_i\Gamma^0_i$, all of which anticommute with $U_F$. One can also consider models satisfying other periodicity, for example $k=3$  \cite{sreejith2016parafermion}, $k=4$ \cite{chew2020time}, or higher \cite{alicea2016topological}, where index $\alpha = 0,\,2\pi/k$ would enumerate oscillating modes $\Gamma_i^\alpha$. 
 
 Despite several advantages such as simplicity and illustrative power, aforementioned models 
 are integrable, and therefore, inherently fine-tuned. A major challenge has been to prove whether they exhibit any robustness towards pertubartions. In this work, we overcome this by deriving the conditions under which, in the presence of perturbations, such integrable models would have oscillating modes with exponentially long lifetimes with respect to the inverse of the perturbation strength. A corollary is that the notion of ``strong'' modes such as Majoranas previously discussed in the static settings \cite{else2017prethermal2}, via our work, is now generalized to nonequilibrium settings. Moreover, we show that disorder can be engineered to stabilize the boundary oscillations.
 %{\color{blue}, rendering Floquet SPT phases.}

\textbf{Model}. -- We study a prototypical example of a driven one-dimensional topological system described by a time-dependent Hamiltonian
\be\label{eq:main_ham}
H(t) =  A(t)\Bigl(J\sum_{i=1}^{L-1} \sigma_i^x\sigma_{i+1}^x+\sum_{i=1}^L h_x^i\sigma^x_i\Bigl)+B(t)h_z\sum_{i=1}^L \sigma^z_i,
\ee
where $\sigma^\alpha_i$ are 2$\times 2$ Pauli matrices, $J$ is a coupling constant, $h^i_x$ and $h_z$ are local fields, and $A(t)=A(t+T)$ and $B(t) = B(t+T)$ are periodic control parameters. 
Below we consider a two-pulse dynamics setting $A(t) = 1$, $B(t) = 0$ for $0\leq t<\tau$, and $A(t) = 0$, $B(t) = 1$ for $t\ge \tau$.  The Hamiltonian in Eq.~\eqref{eq:main_ham} can be implemented in 
superconducting qubit circuits  \cite{levitov2001quantum,you2014encoding,tsomokos2010using}, quantum dots \cite{choy2011majorana,sau2012realizing,fulga2013adaptive}, superfluids \cite{jiang2011majorana,cooper2019topological}, and semiconductor nanowires \cite{stanescu2013majorana}.

\begin{figure}[t!]
    \centering
    \includegraphics[width=0.5\textwidth]{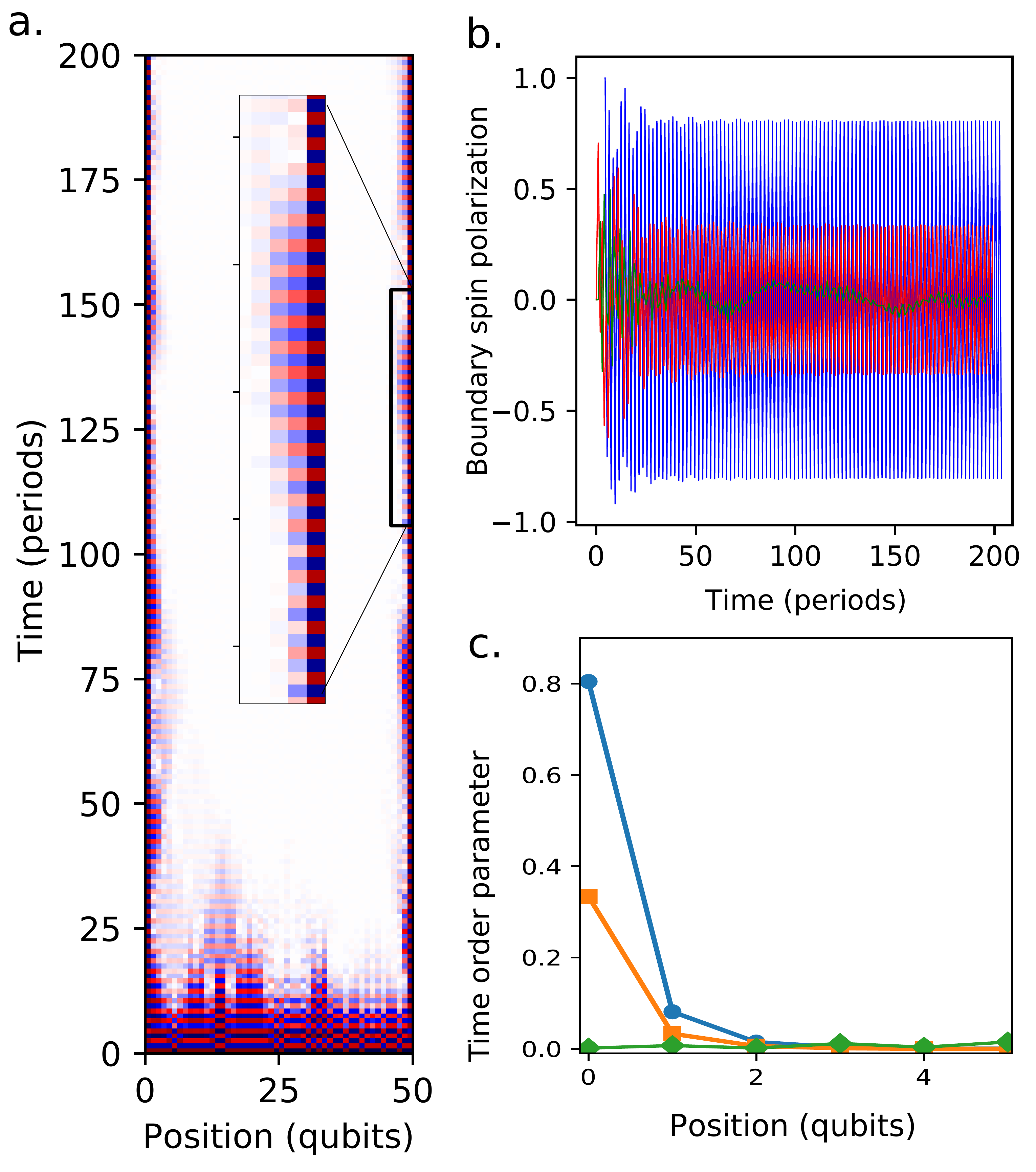}
    \caption{\textbf{Boundary subharmonic response} for the free fermion limit $J\tau = 3\pi/8$, $h_z(T-\tau) = \pi/8$, and $h^x_i=0$. (a) Space-resolved evolution of individual $x$-polarization of $L=50$ qubit chain initiated in a random product state of $x$-polarized qubits. While the bulk oscillations vanish, enlarged inset shows persistent SR oscillations at the boundary. (b) Polarization oscillations for the boundary spin for $x$ (blue), $y$ (red), and $z$ (green) polarizations. (c) Distribution of SR parameter [see Eq.\eqref{eq:order_parameter}] near the boundary in the chain for $x$ (blue), $y$ (orange), and $z$ (green) polarization operators averaged over $N=100$ initial periods.}
    \label{fig:spin_oscillations}
\end{figure}

 To quantify double-period SR for Majorana fermion, one needs a physical parameter that reflects the persistent local oscillations with twice the period. Let 
\be\label{eq:order_parameter}
C(O_\mu) = \frac 12\biggl|\lim_{N\to\infty}\frac1{N}\sum_{n=1}^{N} \Bigl(\<O_\mu(t_{2n})\>-\<O_\mu(t_{2n+1})\>\Bigl)\biggl|
\ee
be the response function that quantifies the periodicity breaking. For numerical and experimental analysis, $C(O_\mu)$ can also be used for finite but large number of periods $N$. The parameter $C$ is the Fourier component of the observable taken at frequency $\omega T= 1/2$ and vanishes for generic systems, while it is nonzero in systems with a subharmonic response \cite{zhang2017j,choi2017observation,rovny2018observation,pal2018temporal}.

\begin{figure*}[t!]
    \centering
    \includegraphics[width=1\textwidth]{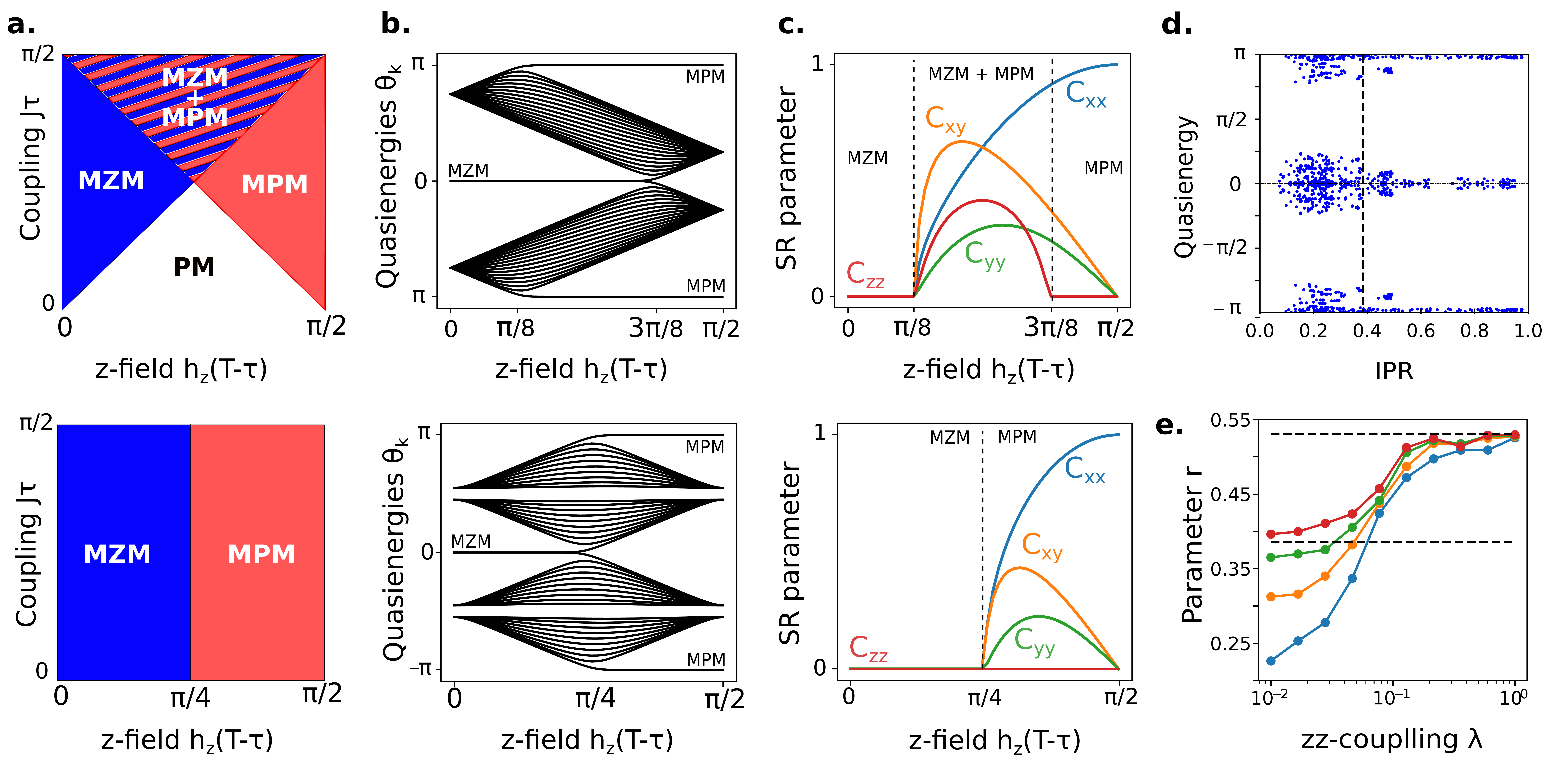}
    \caption{\textbf{Phase diagrams and stability of SR.} (a)-(c) Floquet SPT phases and associated SR parameters for $h_x^i\tau = 0$ (top) and $h_x^i\tau = \pi/2$ (bottom).  (a) Phase diagrams depicting the regions of trivial paramagnetic (PM) and three distinct topological phases exhibiting presence of MZM, MPM, or both. (b) An example of spectra of single fermion quasienergies $\theta_k$ obtained for a system of size $L=20$ spins with parameter $J\tau = 3\pi/8$. (c) The dependence of SR parameters across the phase diagram. The notation $C_{\mu\nu}$ stands for the SR parameter $C(\sigma^\mu_1)$ (see Eq.\eqref{eq:order_parameter}) describing the boundary oscillations of $\mu$ polarization given the initial polarization is chosen in $\nu$ direction. The plots are calculated for a samples of size $L=100$ spins. (d)-(e). Effect of discrete integer-valued disorder and localization. (d) Distribution of single fermion quasienergies and IPR for a model described by Eq. \eqref{eq:corr_static_ham} for $L=500$ and $A=3$. The dots represent individual eigenstates exhibiting strong localization with average IPR $\<\mathcal I_k\>\approx 0.38$. (c) Dependence of level spacing parameter $r$ (defined in the text) as a function of the $zz$-coupling strength $\lambda$ for system sizes $L = 8$ (blue), 10 (orange), 11 (green), and 12 (red) and non-fine-tuned parameters $J\tau= 0.9\cdot\pi/2$, $h_z(T-\tau)= 0.9\cdot\pi/2$. By increasing system size, the curve approaches the crossover curve between  MBL ($r\approx0.386$) and ergodic ($r\approx 0.53$) values with transition at $\lambda T\sim 0.1$. }
    \label{fig:full_picture1}
\end{figure*}

 Let us show how to use a free fermion representation of the Hamiltonian in Eq.$\,$\eqref{eq:main_ham} and construct the Majorana modes. First, consider the Jordan-Wigner transformation \cite{suzuki2012quantum}
$c_i = \frac 12\mathcal P_{z,i} (\sigma^x_i- i\sigma^y_i),$ and  $c^\dag_i = \frac 12\mathcal P_{z,i} (\sigma^x_i+ i\sigma^y_i)$,
where $P_{z,i} = \prod_{k=1}^{i-1}(-\sigma^z_k) $ is a string operator, $c_i$ and $c_i^\dag$ are spinless fermion creation and annihilation operators, $\{c_i,c_j\}= 0$, $\{c^\dag_i,c_j\}= \delta_{ij}$. 
Under the assumption $h^i_x\tau=n\pi/2$, $n\in \mathbb Z$, 
the system is described by a set of stationary orthogonal single-fermion modes
\be\label{eq:sf_modes}
\psi_k(t_{n+1}) = e^{-i\theta_k}\psi_k(t_n),
\ee
where $\psi_k = \sum u_{ki}c_i+v_{ki}c^\dag_i$, $\theta_k\in[-\pi,\pi]$ are single mode quasienergies, $u_{ki}$ and $v_{ki}$ are complex-valued coefficients satisfying the normalization condition $\sum_i|u_{ki}|^2+|v_{ki}|^2 = 1$ (see Supplemental Material). The quasienergy spectrum of single-particle modes is shown in Fig. \ref{fig:full_picture1}(b) for two homogeneous signature cases $h^i_x\tau = 0$ (top) and $h^i_x \tau = \pi/2$ (bottom). In both cases, the Majorana modes $\Gamma^\alpha_i$  can be seen as unique self-adjoint modes with quasienergies $\theta_{k} = 0$ for MZM and $\theta_{k} = \pm\pi$ for MPM confined to the boundaries, i.e., $|u_{k i}|$, $|v_{k i}|$ are evanescent in the bulk and essentially nonzero only at a boundary. The phase diagram in Fig. \ref{fig:full_picture1}(a) shows the appearance of MZM and MPM depending on parameters of the Hamiltonian in Eq.\eqref{eq:main_ham} (see also \cite{khemani2016phase}).

After we introduced the Majorana modes, let us study the double-period oscillations for a local observable operator $O_\mu$. 
For generic initial state $|\Psi\>$ which has exponentially small overlap with any eigenstate of the Floquet operator $U_F$, the expression for SR parameters in Eq.\eqref{eq:order_parameter} is equal to
\be\label{eq:C_majoranas}
\begin{split}
C(O_{\mu}) = \Bigl|\sum_{i}\<\Gamma^\pi_i\>&\tr(\Gamma^\pi_iO_\mu)+\<\mathcal P\Gamma^\pi_i\>\tr(\mathcal P\Gamma^\pi_iO_\mu)\\
&+\sum_{i,j}\<\Gamma^\pi_i\Gamma^0_j\>\tr(\Gamma^\pi_i\Gamma^0_j O_\mu)\Bigl|,
\end{split}
\ee
where  $\<\dots\> = \<\Psi|\dots|\Psi\>$ and $\tr(\dots)$ is normalized trace. 

 Due to the localization of Majorana modes, the double period oscillations are localized at the boundary. Fig.\ref{fig:spin_oscillations}(a) shows the expectation $\<\sigma^x_r\>$ vs the distance to the nearest boundary (denoted by  $r$), and vs. time $t$ for a 1D system initialized in a random product state in the $x$ basis. As can be seen from the long-time dynamics, the order parameter $C(\sigma^x_r)$ is nonvanishing only for the pair of boundary spins  (see Fig.\ref{fig:spin_oscillations}(b) and (c)). This follows from the first two terms of Eq.\eqref{eq:C_majoranas} connecting it to MPM mode. One can initialize a random state in the $z$ basis. In this case, the contribution to $C(\sigma^x_r)$ is given by the third term in Eq.\eqref{eq:C_majoranas} and thus tied to the presence of both MZM and MPM modes. The spatial distribution of this type of oscillations is
\be\label{eq:czr}
C(\sigma^z_r) = \Bigl|\sum_{i}\<\Gamma^\pi_1\Gamma^0_1\>\tr(\Gamma^\pi_1\Gamma^0_1\sigma^z_r)\Bigl|\sim e^{-r/\xi_0-r/\xi_\pi}
\ee
where $\xi_\alpha$ is the confinement length of the $\alpha$ Majorana mode. 
In an experiment, by measuring the profile of  $C(\sigma^z_r)$ one can estimate the harmonic mean of the confinement length of the  MZM ($\alpha=0$) and MPM ($\alpha=\pi$) modes. 
The oscillations can be stronger and more extended if the initial state $|\Psi\rangle$ has a {\it finite} overlap with the eigenstates of Floquet operators (see Supplemental Material for details).

\textbf{Effects of interactions.} -- 
 Strictly speaking, exact strong Majorana modes no longer exist in the presence of generic interactions \cite{else2017prethermal2}. Therefore, the local SR can only be observed within a certain time limited by the timescale of equilibration $\tau_*$ in Floquet systems \cite{else2017prethermal,abanin2017effective,weidinger2017floquet}. In the following theorem we show that there exists a range of parameters for which the equilibriation takes exponential time with respect to inverse interaction strength.

\textbf{Theorem}. \textit{Let $U_F$ be the Floquet operator for the Hamiltonian $H(t) = H(t+T)$, and $\Gamma_i^\alpha$ be unitary operators
%$\mathcal P$ is the parity operator, $\mathcal P^2=1$, 
such that
\begin{enumerate}
    \item $(\Gamma_i^\alpha)^2 = I$ and  $U_F\Gamma^\alpha_i =e^{i\alpha} \Gamma^\alpha_iU_F$,
    \item $\|[O\Gamma_1^\alpha,\Gamma_2^\alpha]\|\leq 2^{-\mu L}$ for arbitrary $\mu>0$, and $|{\rm supp}(U_FO U_F^\dag)| \leq |{\rm supp}(O)|+\Delta$ for arbitrary $\Delta<\infty$, for any operator $O$ with finite connected support,
    \item %{\color{blue} There exists $N\in \mathbb Z$ such that $U_F^{N} =  \mathcal P\Gamma^\alpha_1\Gamma^\alpha_2$}
    There exists $N\in \mathbb Z$ such that $U_F^{N} = \mathcal P\Gamma_1^\alpha\Gamma_2^\alpha$, for certain unitary $\mathcal P$.
\end{enumerate}
Consider the dynamics generated by the new Hamiltonian $H'(t) = H(t) + V(t)$ such that $V(t)=V(t+T)$ is a sum of $S$-local terms, and let  $\eta \equiv S\int_0^T dt||V(t)|| \ll1$ is a small parameter.\\
Then there exists a unitary transformation $\mathcal U$ such that the operators $\tilde \Gamma^\alpha_i = \mathcal U^\dag\Gamma^\alpha_i\mathcal U$ satisfy
\be\label{eq:theorem_result}
\begin{split}
\|\tilde \Gamma^\alpha_i(t_n) -e^{-in\alpha}\tilde \Gamma^\alpha_i\| = O(2^{-c/\eta}\; n),
\end{split}
\ee
when the correction preserves parity $[V(t),\mathcal P]=0$, otherwise, when $[V(t),\mathcal P]\neq0$
\be
\begin{split}
\|\tilde \Gamma^\alpha_i(t_n) -e^{-in\alpha}\tilde \Gamma^\alpha_i\| = O(\eta\; n),
\end{split}
\ee
where $c = [S(2N+3)]^{-1}$ is a constant.
}

Let us briefly analyze the conditions of the theorem whose proof is in Supplemental Material. The first condition just establishes the nature of Majorana operators $\Gamma_\alpha$, as we also described earlier. The second condition ensures that Majorana fermion are localized and spatially separated by the distance $L$. The third condition defines the class of Hamiltonians for which prethermalization of Majorana fermion occurs.

The main result of the theorem  is Eq.~\eqref{eq:theorem_result}, which defines the notion of strong prethermal mode and provides a rigorous proof of existence of prethermal $\tilde \Gamma_i^\alpha$ as approximate integrals of motion for exponentially long times, $\tau_* \sim 2^{c/\eta}T$, and apply to both MZM and MPM. To illustrate this general result, let us consider how it applies to the Hamiltonian in Eq.\eqref{eq:main_ham}. First, let us start from a static limit $h_x = h_z = 0$ and assume $J\tau = \pi/2$. In this case $U_F = \sigma_1^x\sigma_{L}^x = \mathcal P\Gamma_1^0\Gamma_2^0$, $U_F^2=I$. According to the theorem, the system is characterized by the presence of stable MZM $\tilde \Gamma_i^0$. This result is in full agreement with the previous works studying stability of (Floquet) Majorana fermion modes \cite{floquet2017potirniche,else2017prethermal2}. In the driven setting, the result of the theorem provides the evidence of stability of double-period oscillations. For example, consider $h_z(T-\tau)= \pi/2$, $J\tau = \pi/(4m+2)$ and $h_x = 0$. In this case $U^{2m+1}_F = \mathcal \sigma_1^x\sigma_{L}^x=\mathcal P\Gamma_1^\pi\Gamma_2^\pi$. Thus, the system exhibits existence of approximate integrals of motion $\tilde \Gamma_i^\pi$. Although the values of $J\tau$ in this case are fine-tuned, the result may also apply to generic $J$ if we assume that deviations from the fine-tuned case can be incorporated into $V(t)$. As a result, all the region near values $h_z(T-\tau) = \pi/2$ should exhibit stable oscillations.
 
Prethermlaization provides a way to make oscillations long-living but does not extend its lifetime to infinity.  At the same time, the thermalization can be prevented completely by introducing strong local disorder into the system \cite{lazarides2015fate,ponte2015many}. The effect of disorder on discrete time crystals was previously studied in \cite{yao2017discrete}. As shown there, adding strong generic disorder, even if it keeps the system in a topological MBL phase, ``dilutes" the effect of boundary SR due to the presence of bulk oscillations associated with localized states. The distinct boundary oscillations can be preserved by applying a special discrete disorder. Let us set the local $x$ fields by $h_x^i\tau = k_i\pi/2$, where $k_i$ are randomly sampled odd integers,  $k_i\in 2\mathbb Z+1$, $k_i\in[-A,A]$, and $A\geq 3$. This integer-valued disorder has no effect on the initial system without corrections because, for any odd $k_i$, the single spin unitary reduces as $\exp(-i\pi k_i\sigma_i^x/2) = \pm\exp(-i\pi\sigma_i^x/2)$. 
For simplicity, let us choose a particular simple model for the correction, $H'(t)= H(t)+\lambda\sum_i\sigma^z_i\sigma^z_{i+1}$, where $\lambda$ is a small coupling constant, $\lambda T\ll1$. In the absence of the discrete disorder as above, such a term would turn the state of the system into an ergodic phase. Let us illustrate the effect of disorder in the simultaneous limit $J\tau = h_z(T-\tau) = \pi/2$, and $\tau/T\to0$. Neglecting boundary effects, the double period evolution $U^{(2)}_F = \exp(-iH_-)\exp(-iH_+)$, where 
\be\label{eq:corr_static_ham}
H_\pm = \lambda T\sum_i\sigma^z_i\sigma^z_{i+1}\pm\frac\pi{2}\sum_ik_i\sigma^x_i+O(\tau/T)
\ee
This double-period Floquet operator, as the previous one in Eq. \eqref{eq:main_ham}, can be studied using Jordan-Wigner transformation
upon a preliminary transformation $\sigma^x\rightleftarrows \sigma^z$, $\sigma^y\to-\sigma^y$. 
After the mapping to the free fermion modes as in Eq.\eqref{eq:sf_modes}, we study the inverse participation ratio $\mathcal I_k = \sum_i |u_{ki}|^4+|v_{ki}|^4$ for the single fermion modes $\psi_k$ of the Floquet operator in Eq.\eqref{eq:corr_static_ham}. For large system sizes, the values of $\mathcal I_k$ remain finite for finite $\lambda>0$ pointing to strong Anderson localization (see Fig. \ref{fig:full_picture1}(d) for size $L=500$). 
Further, deviation of the parameters $J$ and $h_z$ from the fine-tuned values induces interaction between fermion modes and converts the Anderson localized model into MBL phase. To illustrate the stability of this phase, we study the level spacing parameter $r = \mathbb E[\min(d\Theta_\nu,d\Theta_{\nu+1})/\max(d\Theta_\nu,d\Theta_{\nu+1})]$, where $d\Theta_\nu = \Theta_{\nu+1}-\Theta_\nu$ as function of $\lambda$ at a nonintegrable point, where $\Theta_\mu$ are arguments of eigenvalues of Floquet operator (many-body quasienergies) and the expectation is taken with respect to discrete disorder realizations. Numerical simulations for increasing system sizes point on the existence of regions where $r$ is close to expected localized values $r \approx 0.386$ (see Fig. \ref{fig:full_picture1}(e)). The MBL systems can be considered in the context of prethermalization as the system with thermalization time $\tau_* \to\infty$ preserving the SR oscillations indefinitely long in ideally isolated systems. 

Finally, we address the problem of the presence of the gap protecting the Majorana modes $\tilde \Gamma_\alpha$ from mixing with bulk degrees of freedom as well as suppressing quasiparticle excitations induced by the environment. For weak interactions $\lambda$ the system can be understood in terms of quasiparticle modes $\tilde \psi_k \approx \sum_{k'}\gamma^\lambda_{kk'}\psi_{k'}$ for some unitary $\gamma^\lambda$ depending on $\lambda$. A qualitative random matrix theory analysis \cite{shtanko2018stability,vasilchuk2001law} shows that the transition happens for finite $\lambda \sim \sqrt{G\Delta}$, where $G$ is the quasienergy bandwidth, and $\Delta$ is the gap of noninteracting system (see Supplemental Material). 
 
 \textbf{Discussions.} -- We studied the effect of local unitary subharmonic response (SR) in isolated periodically driven systems. We relate this phenomenon to the existence of unpaired MPM and MZM at the boundaries of 1D topological systems. We have shown the long-living nature of the SR oscillations  in local Hamiltonians and developed a way to protect them using discrete disorder.

 There are several future directions. First, one can use the stable SR in topological quantum computing.  In particular, it was recently shown that  oscillations between MZM and MPM can be used for fault-tolerant quantum memory operations
%as well as for braiding 
using driven 1D $p$-wave superconductors \cite{bauer2019topologically}. 
Another direction would be to generalize the stability theorem to dipolar Hamiltonians with applications to trapped ions \cite{zhang2017j,bernien2017probing,zhang2017observation}, Rydberg atoms \cite{zeiher2016many,endres2016atom}, and nitrogen-vacancy spin impurities in diamond \cite{choi2017observation}. The robustness of the SR oscillations against decohering noise also needs future studies.
 
\begin{acknowledgments}
\textbf{Acknowledgements.} We thank Dominic V. Else, Fangli Liu and Max A. Metlitski for fruitful discussions and suggestions. O.S. acknowledges the support of IBMQ internship program and MIT Energy Initiative fellowship. RM acknowledges the support of the IBM Research Frontiers Institute and funding from the MIT-IBM Watson AI Lab under the project Machine Learning in Hilbert space.
\end{acknowledgments}
 
\bibliography{literature}
\clearpage

\pagebreak

\setcounter{page}{1}
\setcounter{equation}{0}
\setcounter{figure}{0}
\renewcommand{\theequation}{S.\arabic{equation}}
\renewcommand{\thefigure}{S\arabic{figure}}
\renewcommand*{\thepage}{S\arabic{page}}

\onecolumngrid

\begin{center}
{\large \textbf{Supplemental Material for \\``Unitary Subharmonic Response and Floquet Majorana Modes"}}\\
\vspace{0.25cm}
{Oles Shtanko and Ramis Movassagh}
\end{center}
\vspace{1cm}

\twocolumngrid
\subsection{Section 1: Sufficient condition}

In this section we derive the sufficiency condition for subharmonic response to take place in generic systems. We recall that for observable $O_\mu$ to have subharmonic response, is sufficient to prove the existence of the conserved mode unitary operator $\tau_i$ that satisfies 
\be\label{eqs:sufficient_cond}
\tau_i\: U_F= e^{2\pi i/k}\:U_F\:\tau_i\; , \qquad |\tr O_\mu \tau_i|>0.
\ee
 Let $|0n\>$ denote the eigenstates of the operator $\tau_i$ corresponding to the eigenvalues $\tau_i^{0n} = \exp(i\phi_n)$, where $\phi_n$ are phases obeying $\phi_n-\phi_{n'}\neq 2\pi r/k$, $r = 1,\dots k-1$. Then, from the first condition in Eq.~\eqref{eqs:sufficient_cond} it follows that
\be\label{eqs:tau_eignevectors}
|mn\> \equiv U_F^m|0n\>, \quad m=0,\dots k-1,
\ee
are also eigenstates of the operator $\tau_i$ with eigenvalues $\tau_i^{mn} = \exp[i(\phi_n+2\pi m/k)]$. The vectors in Eq.~\eqref{eqs:tau_eignevectors} form the full eigenbasis for both operator $\tau_i$ and $k-$periodic evolution operator $U_F^k$ that commutes with $\tau_i$,
\be\label{eqs:uk_spectral}
U_F^k|mn\> =  \exp(i\Phi_{mn})|mn\>
\ee
where $\Phi_{mn}$ are some phases.

We can rewrite the second condition in Eq.~\eqref{eqs:sufficient_cond} as
\be\label{eqs:sum_of_cyclic_terms}
\begin{split}
0<&|\tr O_\mu\tau_i|^2 \equiv D^{-2}\Bigl|\sum_{n=1}^{D/k}\sum_{m=0}^{k-1} \tau_i^{mn}\<mn|O_\mu|mn\>\Bigl|^2\\
&\leq D^{-2}\Bigl(\sum_{n=1}^{D/k} \Bigl|\sum_{m=0}^{k-1}e^{2\pi im/k}\<mn|O_\mu |mn\>\Bigl|\Bigl)^2\\
&\leq (kD)^{-1}\sum_{n=1}^{D/k} \Bigl|\sum_{m=0}^{k-1}e^{2\pi im/k}\<mn|O_\mu|mn\>\Bigl|^2
\end{split}
\ee
%SOME TRANSOTIION SENTENCE
%\be\label{eqs:sum_of_cyclic_terms}
%(kD)^{-1}\sum_{n=1}^{D/k} \Bigl|\sum_{m=0}^{k-1}e^{-2\pi im/k}\<m'n|U^m_FO_\mu U_F^{\dag m}|m'n\>\Bigl|^2>0
%\ee
where $D$ is the Hilbert space dimension. Here we used triangle inequality to obtain the second line and Cauchy-Schwartz for the third line.

\begin{figure}[t!]
    \centering
    \includegraphics[width=0.45\textwidth]{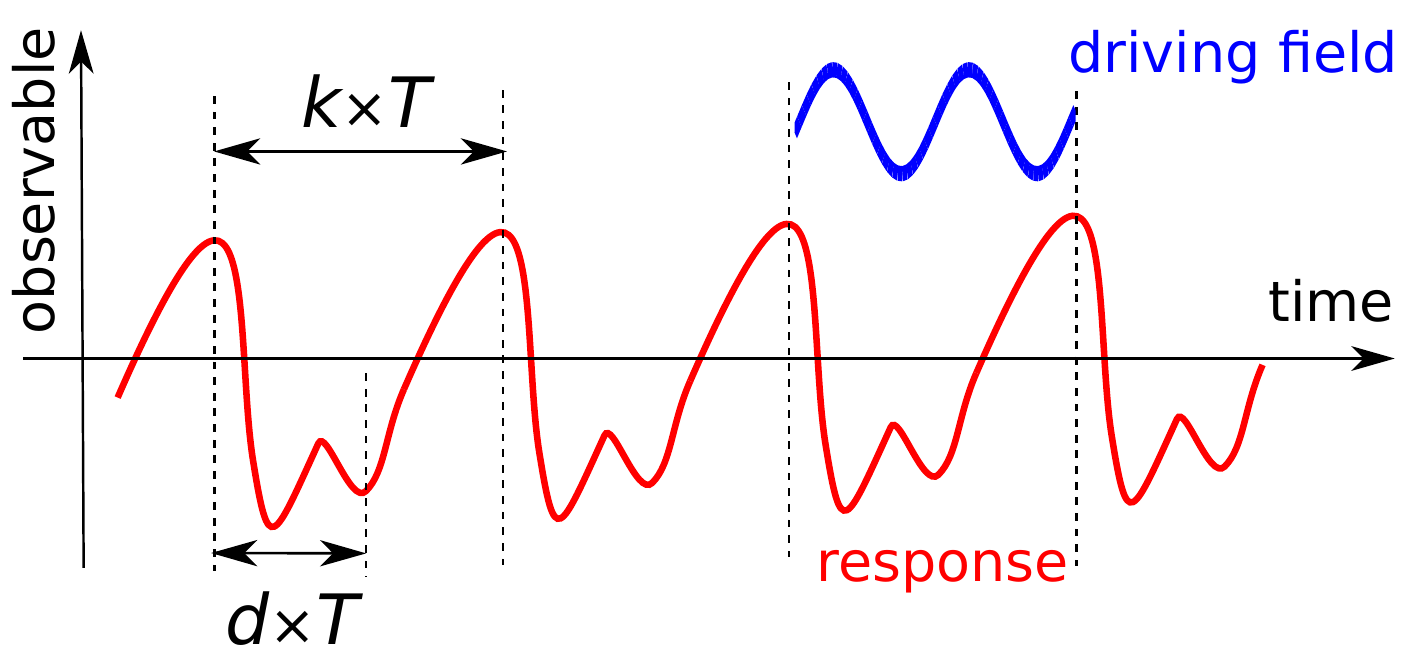}
    \caption{\textbf{Subharmonic response schematics.} The dynamics of observable $\<O_\mu\>$ exhibits dynamics with period $kT$ resulting in $C_k=0$ (see Eq.~\eqref{eqs:sr_parameter}. At the same time, in between the original periodicity of driving field is broken, $C_d>0$ for $0<d\leq k-1$.}
    \label{fig:spin_oscillations}
\end{figure}

Let $S_{osc}$ be the subset of $n\in\{1,\dots,D/k\}$ such that the summand in Eq.\eqref{eqs:sum_of_cyclic_terms} nonzero. $S_{osc}$ contains a constant fraction of all $n$ in the sum. Then for a constant fraction of $n\in S_{osc}$ the RHS of \eqref{eqs:sum_of_cyclic_terms} is nonzero.  From this condition it follows that for some set $S_{\rm osc}$ with non-zero measure, the summands must be nonzero
\be
n\in S_{\rm osc}:\quad \Bigl|\sum_{m=0}^{k-1}e^{2\pi im/k}\<mn|O_\mu|mn\>\Bigl|>0.
\ee
As a result, combining with Eq.~\eqref{eqs:uk_spectral}, for any $n\in S_{\rm osc}$ there exists at least one value of $m$ such that
\be\label{eqs:periodicity}
\exists\; m:\quad \<mn|O_\mu|mn\> = \<mn|U^k_FO_\mu U_F^k|mn\> \neq 0
\ee
Since $\<mn|O_\mu|mn\>=\<m'n|O_\mu|m'n\>$ for all $m$ and $m'$, implies
\be
\begin{split}
\sum_{d=0}^{k-1}e^{2\pi im/k}\<mn|O_\mu|mn\> \propto\sum_{d=0}^{k-1}e^{2\pi im/k} = 0.
\end{split}
\ee
we conclude that at least some matrix elements must be different:
\be\label{eqs:breaking}
\begin{split}
\exists\; m\neq m':\quad \<mn|O_\mu|mn\> \neq& \<m'\,n|O_\mu|m'\,n\>.
\end{split}
\ee

Now let us consider the initial state
\be
|\Psi\> = \sum_{nm}\psi_{nm}|nm\>,
\ee
for some complex amplitudes $\psi_{nm}$.

Let us consider the following quantity to characterize the $k$-periodic subharmonic response,
\be\label{eqs:sr_parameter}
C_d \equiv \Bigl|\lim_{N\to\infty}\frac 1N\sum_{r=0}^N \<U_F^{rk}O_\mu U_F^{\dag rk}\>-\<U_F^{rk+d}O_\mu U_F^{\dag rk+d}\>\Bigl|,
\ee
where $\<\dots\> = \<\Psi|\dots|\Psi\>$. We define subharmonic response to occur if both $C_{d=k}=0$ (that certifies $k$-periodicity of the observable dynamics) and $C_d>0$ for at least one $d\in \{1,\dots,k-1\}$ (that certifies that system breaks the original 1-periodicity), see Fig.\ref{fig:spin_oscillations} for illustration. For $k=2$ the parameter $C_1$ becomes equal to the parameter in Eq.~\eqref{eq:order_parameter} in the main text.
This parameter can be expressed using spectral decomposition as
\be
\begin{split}
C_d &=\Bigl|\lim_{N\to\infty}\frac 1N\sum_{r=1}^N\sum_{nn'=1}^{D/k}\sum_{mm'=0}^{k-1} \exp\Bigl(ir\;\delta\Phi_{nm,n'm'}\Bigl)\\
&\quad \times\psi^*_{nm}\psi_{n'm'}\Bigl(\<mn|O_\mu|m'n'\> - \<mn|U^d_F O_\mu U^{\dag d}_F|m'n'\>\Bigl)\Bigl|.
\end{split}
\ee
where $\delta\Phi_{nm,n'm'} = \Phi_{nm}-\Phi_{n'm'}$ is the difference of phases defined in Eq.~\eqref{eqs:uk_spectral}.
If we assume that the spectrum $\{\Phi_{nm}\}$ has a degeneracy that is $o(N)$, then $\lim_{N\to\infty}N^{-1}\sum_{r=1}^N \exp\Bigl(ir\delta\Phi_{nm,n'm'}\Bigl) = \delta_{nn'}\delta_{mm'}$. As a result, the expression simplifies to
\be
\begin{split}
C_d =\Bigl|\sum_{n\in S_{\rm osc}}\sum_{m=0}^{k-1}|\psi_{nm}|^2&\Bigl(\<mn|O_\mu|mn\>- \<m'n|O_\mu|m'n\>\Bigl)\Bigl|.
\end{split}
\ee
where $m' = (m-d) \,{\rm mod}\, k$.

From Eqs.\eqref{eqs:periodicity} and \eqref{eqs:breaking}, which were derived from the conditions in Eq.\eqref{eqs:sufficient_cond}, it follows that $C_{d=k}=0$, and $C_d>0$ for at least one value of $d\in\{1,\dots, k-1\}$. As a result, the system initialized in a generic state $|\Psi\>$ such that $\sum_{n\in S_{\rm osc}}|\psi_{nm}|^2>0$,  exhibits subharmonic response with the period $kT$.

\begin{figure*}[t!]
    \centering
    \includegraphics[width=1\textwidth]{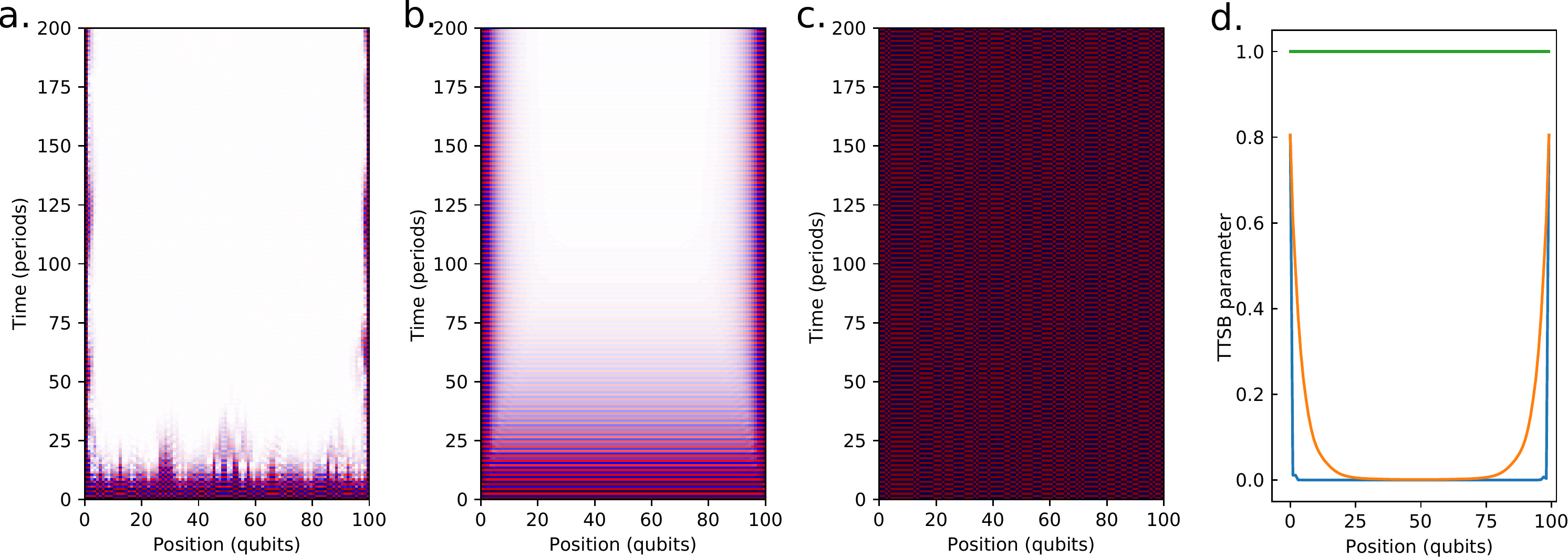}
    \caption{\textbf{Dependence of SR oscillations of fine-tuning of the initial state and parameters of the Hamiltonian.} The system size is $L=100$ qubits. \textit{Panel a}. Evolution of the qubit chain upon initializing the system in a product state with random $x$-polarizations of the qubits. The last term in Eq.\eqref{eq:c_majorana_supp} vanishes in this case, therefore the SR is observed only for first and the last qubit in the chain. \textit{Panel b}. Evolution of the qubit chain upon initializing the system in the product state with the same $x$-orientations. The last term in Eq.\eqref{eq:c_majorana_supp} is non-vanishing for spins close to the boundaries, but vanishes deep into the bulk. This special case shows that SR oscillations happen near the boundary. \textit{Panel c} Evolution of the qubit chain for the fine-tuned case $h_z(T-\tau) = \pi/2$, upon initializing the system in a product state with random $x$-polarizations of the qubits. Any such initial state is a double degenerate eigenstate of the Hamiltonian; therefore the last term in Eq.\eqref{eq:c_majorana_supp} never vanishes for any spin, and the bulk oscillate. This case is a critical point for the transition of the system to DTC. \textit{Panel d}. SR parameter in Eq.\eqref{eq:order_parameter} for Panel a (blue), Panel b (orange), and Panel c (green).}
    \label{fig:osc_profile}
\end{figure*}

\subsection{Section 2: Majorana subharmonic response}

In this section we derive the connection between SR oscillation of local observables and, in particular, the unpaired Majorana modes. First, let us consider the spectral decomposition for the Floquet operator
\be
U_F = \sum_\nu e^{-i\Theta_\nu}|\Phi_\nu\>\<\Phi_\nu|,
\ee
where $\Theta_\nu\in[-\pi,\pi]$ are the many-body quasienergies of the system, and $|\Psi_\nu\>$ are corresponding eigenvectors.

Using the orthogonality of the operators $\Gamma^\pi_i$ and $\Gamma_i^0$, we consider the decomposition
\be
\begin{split}
O_\mu &= \sum_iA^\mu_{i}\Gamma^\pi_1 + \sum_iA^\mu_{i}\mathcal P\Gamma^\pi_i +  \sum_iC^\mu_{i}\Gamma^\pi_i\Gamma^0_i \\
&+\sum_i B^\mu_{i}\Gamma^0_i+\sum_i B^\mu_{i}\mathcal P\Gamma^0_i
+ \tilde O_\mu,
\end{split}
\ee
where $\tilde O_\mu$ represent the rest of the basis decomposition. Then, we can express the local observable operator $O_\mu$ at discrete times $t_n$ as follows
\be\label{eqs:decomp}
\begin{split}
{U_F^\dag}^{n}O_\mu U_F^{n}&  = (-1)^{n} \sum_iA^\mu_{i}\Gamma^\pi_1 + (-1)^{n}\sum_iA^\mu_{i}\mathcal P\Gamma^\pi_i\\
&+(-1)^{n}\sum_iC^\mu_{i}\Gamma^\pi_i\Gamma^0_i+\sum_i B^\mu_{i}\Gamma^0_i+\sum_i B^\mu_{i}\mathcal P\Gamma^0_i\\
&+\sum_{\omega,\nu} e^{-in(\Theta_\omega-\Theta_\nu)}|\Phi_\omega\>\<\Phi_\omega|\tilde O_\mu|\Phi_\nu\>\<\Phi_\nu|.
\end{split}
\ee
Let us focus on the last term and show that its even-times averaged expectation value is
\be\label{eq:odd_tm_av}
\begin{split}
\lim_{N\to\infty}&\frac 1N\sum_{n}\sum_{\omega,\nu} e^{-2in(\Theta_\omega-\Theta_\nu)}\<\Psi|\Phi_\omega\>\<\Phi_\omega|\tilde O_\mu|\Phi_\nu\>\<\Phi_\nu|\Psi\> \\
&= \sum_{\nu,\omega}\delta_{\Theta_\nu-\Theta_\omega,0} \<\Psi|\Phi_\omega\>\<\Phi_\omega|\tilde O_\mu|\Phi_\nu\>\<\Phi_\nu|\Psi\>\\
&\qquad+\sum_{\nu,\omega}\delta_{|\Theta_\nu-\Theta_\omega|,\pi} \<\Psi|\Phi_\omega\>\<\Phi_\omega|\tilde O_\mu|\Phi_\nu\>\<\Phi_\nu|\Psi\>,
\end{split}
\ee
where we used the identity 
\be
\lim_{N\to\infty}\frac 1N \sum_{n=1}^{N} e^{-inx} = \delta_{x,0}+(-1)^n\delta_{|x|,\pi},
\ee
and $\delta_{a,b}$ is a Kronecker delta. 

One may compare the expression in Eq.\eqref{eq:odd_tm_av} with the odd-times average expectation value which differs by the sign of the second term,
\be\label{eq:evn_tm_av}
\begin{split}
\lim_{N\to\infty}&\frac 1N\sum_{n}\sum_{\omega,\nu} e^{-i(2n+1)(\Theta_\omega-\Theta_\nu)}\<\Psi|\Phi_\omega\>\<\Phi_\omega|\tilde O_\mu|\Phi_\nu\>\<\Phi_\nu|\Psi\> \\
&= \sum_{\nu,\omega}\delta_{\Theta_\nu-\Theta_\omega,0} \<\Psi|\Phi_\omega\>\<\Phi_\omega|\tilde O_\mu|\Phi_\nu\>\<\Phi_\nu|\Psi\>\\
&\qquad-\sum_{\nu,\omega}\delta_{|\Theta_\nu-\Theta_\omega|,\pi} \<\Psi|\Phi_\omega\>\<\Phi_\omega|\tilde O_\mu|\Phi_\nu\>\<\Phi_\nu|\Psi\>
\end{split}
\ee
Using the the orthogonality condition, one may express $A^\mu_{\alpha} = \Tr(\Gamma_i^\pi O_\mu)$ and $ B^\mu_{\alpha\beta} = \Tr(\Gamma^\pi_i\Gamma^0_j O_\mu)$. Combining this result with Eq.\eqref{eq:odd_tm_av} and Eq.\eqref{eq:evn_tm_av}, we arrive at the expression for SR that we used in the main text
\be\label{eq:c_majorana_supp}
\begin{split}
C_\mu &=\frac 1{N} \Bigl|\sum_{i}\<\Gamma^\pi_i\>\Tr(\Gamma^\pi_iO_\mu)+\<\mathcal P\Gamma^\pi_i\>\Tr(\mathcal P\Gamma^\pi_iO_\mu)\\
&+\sum_{i,j}\<\Gamma^\pi_i\Gamma^0_j\>\Tr(\Gamma^\pi_i\Gamma^0_j O_\mu)\\
&+\sum_{\nu,\omega}\delta_{|\Theta_\nu-\Theta_\omega|,\pi} \<\Psi|\Phi_\omega\>\<\Phi_\omega|\tilde O_\mu|\Phi_\nu\>\<\Phi_\nu|\Psi\>\Bigl|
\end{split}
\ee
where  $\<\dots\> = \<\Psi|\dots|\Psi\>$.

We now study the role of the last term in Eq.\eqref{eq:c_majorana_supp}. First, we focus on the case where this term is non-negligible for all spins, for example $h_z(T-\tau) = \pi/2$ and $|\Psi\> =2^{-L/2} |+\>$, where we use a notation $|+\> = (|0\>+|1\>)^{\otimes L}$. Then, for any $J$ the Floquet Hamiltonian has a pair of eigenstates $|\Phi_1\> = 2^{-L/2}(|+\>+|-\>)$ with quasienergy $\theta_1 = 0$ and $|\Phi_2\> = 2^{-L/2}(|+\>-|-\>)$ with quasienergy $\theta_1 = \pi$. Simultaneously, $|\Phi_1\>$ and $|\Phi_2\>$ are the only eigenstates with non-zero overlaps with $|\Psi\>$. Let us consider the observable $\sigma_x^i$, $i\neq 1,L$. The SR for these observables is given by the last term equal to 1. Therefore, all spins oscillate without a decay (see Fig. \ref{fig:osc_profile}c). 

The intermediate case is possible if the parameters are not fine-tuned but $|\Psi\>$ is a homogeneous states, e.g., $|\Psi\> = 2^{-L/2}|+\>$ as above.
In this case the oscillations decay into the bulk with a characteristic lengths much larger that Majorana fermion lengtscale $\xi_\alpha$ (see Fig. \ref{fig:osc_profile}b).

Finally, if we assume that $|\<\Psi|\Phi_\nu\>|^2\sim 2^{-L}$, the last term  has exponentially vanishing contribution for in Eq.\eqref{eq:c_majorana_supp}. Therefore, the oscillations in the bulk vanish (see Fig. \ref{fig:osc_profile}a).

\subsection{Section 3: Proof of the main theorem}

For each Hermitian operator $O(t)$ we consider a decomposition
\be
O(t) =\sum_{\alpha=1}^{2^L}\sum_{i=1}^L \frac 1{S_\alpha}\;\xi_{i,\alpha}(t)\; P_\alpha
\ee
where $\xi_{i,\alpha} = s_{i,\alpha}\Tr [O(t)P_\alpha]$ are real-valued coefficients, $P_\alpha$ are generalized Pauli matrices, $s_{i,\alpha} = 1$ if $i\in{\rm supp}(P_\alpha)$ and $s_{i,\alpha}=0$ otherwise, and  $S_\alpha\equiv |{\rm supp}(P_\alpha)|$ is the size of the support of operator $P_\alpha$. The normalization is such that
\be
\sum_{\alpha}\frac 1{S^2_\alpha}\bigl(\sum_i\xi_{i,\alpha}(t)\bigl)^2 =\frac 1{2^L} \Tr[ O^2(t)]
\ee
Then, we define a parametrized family of norms
\be
||O||_\kappa =  \sup_i\sum_\alpha  \overline{|\xi_{i,\alpha}(t)|}\; e^{\kappa S_\alpha},
\ee
where $\kappa >0$ is a real parameter and $\overline {x(t)}$ denotes the time average of the function $x(t)$ over period $T$. We refer to operators $\|O\|_\kappa<\infty$ for some $\kappa>0$ as quasi-local operators.\\

We employ the following notation
\be
\|O\|_n  \equiv  \|O\|_{\kappa_n}, \qquad \kappa_n = \frac{\kappa_0}{1+\log (n+1)}
\ee
for some $\kappa_0>0$ and the positive integer $n$.

We also define a $2N$-dimensional group $\mathcal G_{2N,\Delta} =\{X^k\}$ of local unitary transformations generated by a unitary
\be
\label{eq:Xgroup}
X^{2N} = I, \qquad |{\rm supp}(XO X^\dag)| \leq |{\rm supp}(O)|+\Delta
\ee
where $O$ is an operator with connected finite support. Under the condition of the theorem, we have $U_F \in \mathcal G_{2N,\Delta}$ for some finite $\Delta$.

Consider the full Floquet operator
\be
U'_F =\mathcal T \exp\biggl(-i\int_0^T (H(t)+V(t))dt\biggl),\\
\ee

Let us prove the following theorem that connects $U'_F$ and $U_F$; it serves as a generalization of Theorem 1 in Ref.~\cite{else2017prethermal}.\\

\textbf{Theorem S1}. \textit{Assume $U_F \in \mathcal G_{2N,\Delta}$ and $V(t)$ satisfies
$\eta = \|V\|_{\kappa_0}T/\kappa_0\ll1$ for some $\kappa_0<\infty$. Then there exists a unitary operator $\mathcal U$ such that
\be
\mathcal U \;U'_F\; \mathcal U^\dag = U_F\; U_{\rm corr}
\ee
where
\be
\label{eq:Y_unitary}
U_{\rm corr} = \mathcal T \exp\Bigl(-i\int_0^T\Bigl(D+\mathcal V(t)\Bigl)dt\Bigl)
\ee
satisfying $[D,U_F] = 0$ and
\be
\frac{\|D\|_{n_*}}{\|V\|_0}\leq 2e^{2\kappa_0\Delta N}, \quad \frac{||\mathcal V||_{n_*}}{||V||_0}\leq O(2^{-n_*})
\ee
where $n_* = O\Bigl(\kappa_0/2\eta(N+3)\Bigl)$. }\\

%Additionally,
%\be
%\|\mathcal U\|_{\kappa_{n_*}}\leq \exp(c\|V\|_0)  
%\ee
%i.e. it is a quasi-local operator.}\\

\textbf{Theorem S2}.(Abanin, De Roeck, Ho, Huveneers \cite{abanin2017rigorous})\textit{ Consider the operator $O$ that has a finite support $S$ and unitary transformation in the form $U_{\rm corr}$ in Eq.\eqref{eq:Y_unitary} such that $||D||_\kappa<\infty$ and $||\mathcal V||_{\kappa'}\ll 1$ for some $\kappa$ and $\kappa'$. Then
\be
|| U^{\dag} _{\rm corr}\; O\; U_{\rm corr}  - e^{-iDT}\:O\:e^{iDT}|| \leq c_1\|O\|\|\mathcal V\|_\kappa(T+c_2)
\ee\
for some $c_1>0$ and $c_2>0$ independent of $T$.
}\\

The proof of Theorem S1 is provided below in this section and is a $\Delta>0$ generalization of Theorem 1 from Ref.~\cite{else2017prethermal}. First let us note that
\be
\begin{split}
|| (U_F')^{\dag n}& \tilde \Gamma^\alpha_i  (U_F')^{n}   - e^{-in\alpha} \tilde \Gamma^\alpha_i ||\\
&\leq \sum_{k=1}^{n}\Bigl\|e^{i(k+1)\alpha}(U_F')^{\dag k}  \tilde \Gamma^\alpha_i  (U_F')^{k} \\
&\qquad\quad -\-e^{ik\alpha}(U_F')^{\dag {k-1}} \tilde \Gamma^\alpha_i   (U_F')^{k-1}\Bigl\| \\
& = n\|(U_F')^{\dag}   \tilde \Gamma^\alpha_i   U'_F - e^{-i\alpha} \tilde \Gamma^\alpha_i \|\\
& = n ||U_{\rm corr}^\dag U_F^{\dag}\Gamma^\alpha_i  U_F U_{\rm corr}- e^{-i\alpha}\Gamma^\alpha_i || \\
& = n ||U_{\rm corr}^\dag\Gamma^\alpha_i  U_{\rm corr}- \Gamma^\alpha_i || 
\end{split}
\ee

%\be
%\begin{split}
%||(U_F)^n&  \tilde \Gamma_\alpha  (U^\dag_F)^n - e^{in\alpha} \tilde \Gamma_\alpha|| \\
%&=||(U_F)^n \mathcal U \Gamma_\alpha \mathcal U^\dag (U^\dag_F)^n - e^{in\alpha}\mathcal U\Gamma_\alpha\mathcal U^\dag||  \\
%&=||(Y^\dag U_0^{\dag})^n\Gamma_\alpha (U_0 Y)^n- e^{in\alpha}\Gamma_\alpha|| \\ &=
%||\left(\prod_{k=0}^n Y_k\right)^\dag \Gamma_\alpha  \prod_{k=0}^n Y_k^n- \Gamma_\alpha||
%\end{split}
%\ee

Using Cauchy-Schwarz inequality, we obtain
\be
\label{eq:decomp_GG}
\begin{split}
||U_{\rm corr}^\dag\Gamma^\alpha_i  U_{\rm corr}- \Gamma^\alpha_i || &\leq ||U_{\rm corr}^\dag\Gamma^\alpha_i  U_{\rm corr}- e^{-iDT}\Gamma_\alpha e^{iDT}||\\&+
||e^{-iDT}\Gamma^\alpha_i  e^{iDT}-\Gamma^\alpha_i||
\end{split}
\ee
If $[V(t), \mathcal P]=0$, then also $[D,\mathcal P] = 0$. Because $[D,U_F]=0$, therefore $[D,U^N_F]=0$ and
\be
[D,U^N_F] = [D,\Gamma_1^\alpha\Gamma^\alpha_2] = \Gamma_1^\alpha[D,\Gamma^\alpha_2] + [D,\Gamma_1^\alpha]\Gamma^\alpha_2 = 0. 
\ee
From this expression we find
\be
[D,\Gamma_1^\alpha] = \Gamma_2^\alpha [ D\Gamma_1^\alpha, \Gamma_2^\alpha].
\ee
Then according to the condition 2 of the theorem,
\be
\|[D,\Gamma_1^\alpha]\| \leq \|[ D\Gamma_1^\alpha, \Gamma_2^\alpha]\|\leq e^{-\kappa L}\to 0 
\ee
 where $\kappa = \max(\kappa_{n_*},\mu)$.

As a result, we get the expression
\be
\|(U_F)^n  \tilde \Gamma_\alpha (U^\dag_F)^n - e^{i\alpha} \tilde \Gamma_\alpha\|\leq O(2^{-n_*} n)
\ee
where $n_*$ is given by Theorem S1. If the operator $V(t)$ has finite support $S$, then $\|V\|_\kappa \leq e^{\kappa S}$. In this case $\|V\|_\kappa/\kappa$ has minimum at $\kappa = S^{-1}$.  This proves the first part of the main theorem in the main text.

If $[V(t),\mathcal P_z]\neq0$, the last term in Eq.\eqref{eq:decomp_GG} is dominant, therefore
\be
||U_{\rm corr}^\dag\Gamma^\alpha_i  U_{\rm corr}- \Gamma^\alpha_i ||\leq O(\|D\|_{n_*})
\ee
Using Theorem S1, we can also bound this expression.
\\

%If $D$ \textit{does not} respect the $\mathbb Z_2$ symmetry, then a less tight bound is applied
%\be
%||e^{-iDT}\Gamma^\alpha_i  e^{iDT}-\Gamma^\alpha_i||\leq 2e^{\kappa_0\Delta}\|C\|_{\kappa_0}T
%\ee
% |[H,AB]| < |[H,A]|+|[H,B]|
%which leads to
%\be
%\|(U_F)^n  \tilde \Gamma_\alpha (U^\dag_F)^n - e^{i\alpha} \tilde \Gamma_\alpha\|\leq O(n\|C\|_{\kappa_0}/\omega)
%\ee
%which leads to the statement of the theorem. \\

\textit{Proof of Theorem S1}. 
%We define the average value
%\be
%\<O\>_X = \frac 1N \sum_{k=0}^{N-1}X^k O X^{-k}
%\ee
%By construction, $[\<O\>_X,X]=0$. 
Following steps from Theorem 1 in Ref.~\cite{else2017prethermal} we construct a sequence of operators $U_n$ such that
\be
U_{n+1} = \mathcal U^\dag_n\; U_0\; \mathcal U_n,\qquad \mathcal U_n = \prod_{k=0}^{n-1} e^{iA_k},
\ee
where $U_0\equiv U'_F$ and $A_k$ are Hermitian operators we define below. For each $U_n$ we consider a decomposition
\be
U_n = U_F\mathcal T \exp\Bigl(-i\int_0^T H_n(t)\;dt\Bigl)
\ee
for a non-unique choice of the time-dependent operator $H_n(t)$.
Our goal is to show the existence of an \textit{optimal} choice for the sequence $A_n$ and the operators $H_n(t) = D_n + \mathcal V_n(t)$, such that
\be
D_n = \<\overline{H_n}\>_{U_F} \equiv \frac 1{2N} \sum_{k=0}^{2N-1}U_F^k\overline{H_n} U_F^{-k} = 0
\ee
where overbar denotes the time average, and the norm of operator $\mathcal V_n$ decreases exponentially with $n$ if $n\leq n_*$.

Assume that we found the sequence $H_k(t)$ for $k\leq n$. Let us show the procedure for $H_{n+1}(t)$. For this, we rewrite
\be
\label{eq:floquet_suboptimal}
\begin{split}
U_{n+1} & = e^{-iA_n}U_n e^{iA_n}\\
&=U_F \Bigl[U_F^\dag e^{-iA_n} U_F \mathcal T \exp\Bigl(-i\int_0^TH_n(t)\;dt\Bigl) e^{iA_n}\Bigl] \\&=
\mathcal T \exp\Bigl(-i\int_0^TH'_{n+1}(t)\;dt\Bigl)
\end{split}
\ee
where $H'_{n+1}(t)$ represents a (suboptimal) decomposition which easily follows from Eq.\eqref{eq:floquet_suboptimal} by
\be
H'_{n+1}(t) = 
\begin{cases}
\tau^{-1}A_n,\qquad &0<t\leq\tau \\
(1-2\tau/T)^{-1}H_n(t'), \qquad &\tau<t\leq T-\tau\\
-\tau^{-1}U_F^\dag A_nU_F, \qquad &T-\tau <t\leq T
\end{cases}
\ee
where $0<\tau<T/2$ is an arbitrary real parameter and $t' = T(t-\tau)/(T-2\tau)$.

First, let us decompose the correction time-dependent potential into static and zero-average components,
\be
\mathcal V_n(t) = E_n + \delta \mathcal V_n(t)
\ee
such that $\overline{\delta \mathcal V_n}=0$. Then the time-averaged value of the Hamiltonian is
\be
\overline{H'}_{n+1} = D_n + E_n+A_n-U_F^\dag A_nU_F
\ee
The time-dependent part is bounded as
\be
\begin{split}
||\delta \mathcal V'_{n+1}||_{\kappa_n} &= ||H'_{n+1}-\overline{H'_{n+1}}||_{\kappa_n} \\
&\leq  2||A_n||_{\kappa_n} +||E_n||_{\kappa_n}+||\delta \mathcal V_n||_{\kappa_n} +4\tau ||D_n||_{\kappa_n} 
\end{split}
\ee
Following Ref.~\cite{else2017prethermal}, we choose
\be
A_n = \frac 1{2N} \sum_{k=0}^{2N-1}\sum_{p=0}^{k}U_F^pE_nU_F^{-p}
\ee
With this choice
\be
\overline{H'_{n+1}} = D_n, \qquad \|A_n\|_\kappa\leq \frac 12\alpha(\kappa) \|E_n\|_\kappa
\ee
where $\alpha(\kappa) = (2N+1)e^{2\kappa N\Delta}$.

As a result,
\be
\label{eq:dV_bound1}
||\delta \mathcal V'_{n+1}||_{\kappa} \leq  (\alpha(\kappa) +1)||E_n||_{\kappa} + ||\delta V_n||_{\kappa}
\ee
According to Theorem 1 in Ref.~\cite{abanin2017rigorous}, there exists a unitary  $Y(t) = Y(t+T)$, with $Y(0) = I$ such that
\be
\label{eq:unitaries_adjustment}
H_{n+1}(t) = Y_n(t) H'_{n+1}(t) Y^\dag_n(t) - iY_n(t)\partial_tY^\dag_n(t)
\ee
and, under the condition $3||\delta \mathcal V_n||_n\leq \kappa_n-\kappa_{n+1}$, the transformed Hamiltonian satisfy
\be
\label{eq:pretherm_bounds}
\begin{split}
&||\overline{H_{n+1}}-\overline{H'_{n+1}}||_{n+1}\leq {\epsilon_n}/2\\
&||\delta \mathcal V_{n+1}||_{n+1} \leq {\epsilon_n}
\end{split}
\ee
where
\be
\label{e_bound}
\epsilon_n = Tm_n ||\delta \mathcal V'_{n+1}||_n\Bigl(||\overline {H'_{n+1}}||_n+2||\delta \mathcal V'_{n+1}||_n\Bigl)
\ee
and
\be
m_n = \frac{18}{\kappa_{n+1}(\kappa_n-\kappa_{n+1})}
\ee

Using this result, we obtain the optimal Hamiltonian $H_{n+1}(t)$ from suboptimal $H_{n+1}'(t)$. The parameters of optimal Hamiltonian satisfy the following bounds
\be
\label{eq:deltaD}
\begin{split}
\|D_{n+1}-D_n\|_{n+1} = \|\<D_{n+1}+E_{n+1}-D_n\>_{U_F}\|_{n+1}\\
\leq \beta(\kappa_{n+1}) \epsilon_n/2
\end{split}
\ee
where $\beta(\kappa) = e^{2N\kappa \Delta}$. Also
\be
\label{eq:E_bound}
\begin{split}
\|E_{n+1}\|_{n+1}&\leq \|D_{n+1}+E_{n+1}-D_n\|_{n+1}\\
&\qquad+\|D_{n+1}-D_n\|_{n+1}\leq \gamma(\kappa_{n+1})\epsilon_n
\end{split}
\ee
where $\gamma(\kappa) = (1+\beta(\kappa))/2$.

Now, let us use the induction. Assume that for $n$th step the operatos obey
\be
\label{eq:iterative_bounds}
\begin{split}
&||E_n||_{n}\leq 2^{-n}\gamma(\kappa_n)\lambda\\
&||\delta \mathcal V_n||_{n}\leq 2^{-n}\lambda
\end{split}
\ee
as well as
\be\label{eq:iterative_bounds2}
||D_{n+1}-D_n||_{n+1}\leq 2^{-n-1}\beta(\kappa_{n+1})\lambda
\ee
where we denote $\lambda = 2\|V\|_0$.

First, we need to verify Eq. \eqref{eq:iterative_bounds} for $n=0$. Let us set
\be
U_0 = U'_F = U_F\exp\left(-i\int_0^T  dt H_0(t)\right),
\ee
where $H_0(t) = U_t V(t)U_t^\dag$ and
\be
U_t = \mathcal T\exp \left(-i\int_0^t dt' H(t')\right).
\ee
We derive $\|D_0\|_0 = \|\<\overline{H_0}\>_{U_0}\|_0\leq\beta(\kappa_0)\lambda/2$ as well as
\be
\begin{split}
&\|E_0\|_0 = \|\overline{H_0}-D_0\|_0 \leq \gamma(\kappa_0)\lambda\\
&\|\delta V_0\|_0 = \|H_0-\overline {H_0}\|_0\leq \lambda
\end{split}
\ee

Now, for $n\geq1$ we substitute Eq.\eqref{eq:iterative_bounds} into Eq.\eqref{eq:dV_bound1} and, in turn, using this expression in
%\be
%||\delta V'_{n+1}||_{n}\leq 2^{-n+1}(\alpha(\kappa_{n})+1)\lambda
%\ee
Eq.\eqref{e_bound} leads to
\be
\epsilon_n \leq 2^{-n} \xi_n m_n \lambda T  ||D_n||_{n}%+O(4^{-n})
\ee
where $\xi_n \equiv (\alpha(\kappa_{n})+1)\gamma(\kappa_n)+1$.

Using Cauchy-Schwarz inequality, we can estimate that
\be
\begin{split}
\|D_n\|_{n}& =  \Bigl\|D_0+\sum_{k=0}^{n-1} D_{k+1}-D_k\Bigl\|_{n}\\
&\leq \|D_0\|_n+\sum_{k=0}^{n-1}||D_{k+1}-D_k||_n\\
&\leq \beta(\kappa_0)\lambda/2+\sum_{k=0}^{n-1}||D_{k+1}-D_k||_{k+1}\\
&\leq \beta(\kappa_0)\lambda+O(2^{-n}\lambda)
\end{split}
\ee
where we used that for $m<n$ norms satisfy $||O||_{m}\geq \|O\|_n$ as well as the bounds from Eq.\eqref{eq:iterative_bounds2}. As the result we obtain
\be
\epsilon_n \leq 2^{-n}\lambda^2 T \xi_nm_n\beta(\kappa_0)+O(4^{-n})
\ee
Taking into account Eqs.\eqref{eq:pretherm_bounds}, \eqref{eq:deltaD}, and \eqref{eq:E_bound}, the step $n+1$ is satisfied if
\be
\xi_nm_n\beta(\kappa_0)\lambda T\leq \frac 12
\ee
Assuming that $\lambda T\ll 1$, this expression is valid for $n\leq n_*$, where 
\be
n\leq n_* = O\Bigl(\kappa_0^2/(2N+3)\lambda T\Bigl).
\ee
Because $||\delta \mathcal V_0||_n\leq ||\delta \mathcal V_n||_0 = \lambda$, the condition $3||\delta \mathcal V_n||_n\leq \kappa_n-\kappa_{n+1}$ (see paragraph after Eq.\eqref{eq:unitaries_adjustment}) is satisfied for the conditions of the theorem, $\lambda T/\kappa_0\ll 1$.

Denoting $\mathcal U \equiv \mathcal U_{n_*}$, $D \equiv  D_{n_*}$ and $V(t) \equiv  E_{n_*}+\delta \mathcal V_{n_*}$, we prove statement of the theorem.

\subsection{Section 4: Free fermion solution (Fig. \ref{fig:spin_oscillations} and Fig. \ref{fig:full_picture1}a-c)}
Let us consider the time-periodic Hamiltonian in Eq.\eqref{eq:main_ham} in the main text for discrete $x$-field values,
\be
H(t) =  A(t)\Bigl(J\sum_i \sigma_i^x\sigma_{i+1}^x+\frac \pi{2\tau}\sum_ik_i\sigma^x_i\Bigl)+B(t)h_z\sum_i \sigma^z_i
\ee
where $k_i \in \mathbb Z$ are integer variables, $A(t) = 1$, $B(t) = 0$ for $0\leq t<\tau$, and $A(t) = 0$, $B(t) = 1$ for $t\geq\tau$.

The Floquet Hamiltonian corresponding to the Hamiltonian $H(t)$ is
\be
U_F = U_ZU_XU_P,
\ee
where the unitary operators are defined as follows
\be
\begin{split}
&U_Z= \exp\Bigl(-iJ\tau\sum_i \sigma_i^z\Bigl),\\
&U_X = \exp\Bigl(-ih_z(T-\tau)\sum_i \sigma_i^x\sigma_{i+1}^x\Bigl),\\
&U_P=\prod_i (\sigma^x_i)^{[k_i/2]}
\end{split}
\ee
where $[x/2]$ is modulo operation acting on integer $x$, it returns $0$ if $x$ is even and $1$ if $x$ is odd. 

First, the summands in the Hamiltonian are quadratic in the fermion operators
\be
\begin{split}
&\sum_i \sigma_i^z = \sum_i c_i^\dag c_i - c_i c^\dag_i,\\
&\sum_i \sigma_i^x\sigma_{i+1}^x = \frac 12\sum_i c_i^\dag c_{i+1}+c^\dag_{i+1}c_i + c_{i+1}^\dag c^\dag_i+ c_ic_{i+1}
\end{split}
\ee

Also, the action of the unitary operator $U_P$ is
\be
\begin{split}
U_Pc_m&U_P^\dag = \prod_{i=1}^L (\sigma^x_i)^{[k_i/2]}\prod_k^{m-1} (-\sigma^z_k)\sigma^-_m\prod_{j=1}^L (\sigma^x_{j})^{[k_j/2]}\\
&= \prod_{i=1}^{m-1} (\sigma^x_i)^{[k_i/2]}\prod_k^{m-1} (-\sigma^z_k)\prod_{j=1}^{m-1} (\sigma^x_{j})^{[k_j/2]}\sigma^x_m\sigma^-_m\sigma^x_m \\
&= (-1)^{R_m}c_m^\dag
\end{split}
\ee
where $R_m = \sum_i^{m-1}[k_i/2]$.

Let us introduce the vector $\psi = (c_1\dots c_L,c_1^\dag\dots c_L^\dag$). The resulting transformation can be written as
\be\label{eq:single_fermion_tranform}
U_F\;\psi_i\;U_F^\dag = \sum_i V_{ij}\psi_j
\ee
where the $2L\times 2L$ unitary matrix $V$ is a \textit{single-fermion Floquet operator}. For the problem above, it can be presented in the form of the product
\be\label{eqs:single_particle_u}
V = V_ZV_XV_P
\ee
where
\be
\begin{split}
&V_Z = \exp\Bigl(-i\theta_1\Bigl[(\tau_z+i\tau_y)\sum_i |i\>\<i+1|+{\rm h.c.}\Bigl]\Bigl),\\
&V_X = \exp\Bigl(-2i\theta_2\tau_z\sum_i |i\>\<i|\Bigl), \\
&P = \sum (-1)^{R_i}\tau^{[k_i/2]}_x|i\>\<i|, \qquad R_i = \sum_{j=1}^{i-1}[\xi_j/2].
\end{split}
\ee
where $\tau_i$ are Pauli matrices associated to creation and annihilation operators (Nambu space). The $V$ has eigenvalues $\exp(\pm i\theta_k)$ and corresponding eigenvectors $\psi_k,\psi'_k = (u_{ki},v_{ki}),(v^*_{ki},u^*_{ki})$ setting the free fermion representation discussed in the main text. The spectrum calculated using operators $V$ in Eq.\eqref{eqs:single_particle_u} is shown in Fig. \ref{fig:full_picture1}b in the main text.

The system dynamics can be characterized by one- and two-point correlator functions,
\be\label{eq:phi_spdm}
\phi_i(t) = \<\Psi_t|\psi_i|\Psi_t\>,\qquad \rho_{ij}(t) = \<\Psi_t| \psi^\dag_i\psi_j|\Psi_t\>
\ee
where we call $\phi(t)$ a vector of operator expectations  and $\rho(t)$ a single-particle density matrix. 

The evolution of the correlators in Eq. \eqref{eq:phi_spdm} is given by
\be
\phi(t_{n+1}) = V\phi(t_n), \qquad \rho(t_{n+1}) = V\rho(t_n)V^\dag
\ee
To evaluate the evolution of $\phi(t_{n})$ and $\rho(t_{n})$ using the equations above, we need to know the initial conditions, $\phi(0)$ and $\rho(0)$. Below, we provide the initial conditions for several relevant spin configurations.

Assume initially all the qubits are polarized in $x$-direction, $|\Psi\> = \Motimes_{i=1}^L|s_i\>_x$, where the coefficients $s_i = \pm 1$ represent a binary vector and $|k\>_\alpha$ are eigenvalues of the operator $\sigma^\alpha$ with corresponding eigenvectors $k=\pm 1$. Then, the initial values of the correlators are
\be
\begin{split}
&\phi(0)=|\phi_x \> = (\frac {s_1}2,0,\dots,\frac {s_1}2,0,\dots),\\
&\rho(0) = \rho_x(\vec s) \equiv 
\frac 14\left(\begin{matrix}
{\rm diag}_{3}(\vec d,2,\vec d)&{\rm diag}_{3}(-\vec d,0,\vec d)\\
{\rm diag}_{3}(\vec d,0,-\vec d)& {\rm diag}_{3}(-\vec d,2,-\vec d)
\end{matrix}\right)
\end{split}
\ee
where $d_i = s_{i+1}-s_i$, ${\rm diag}(\vec x)$ is a diagonal matrix with elements $x_i$ on the diagonal, ${\rm diag}_3(\vec x,n,\vec y)$ is a tridiagonal matrix with all diagonal elements equal to $n$, and $x_i$ and $y_i$ on lower and upper diagonals respectively.

Similar expression can be obtained for the product of $y$-spins, $|\Psi\> = \Motimes_{i=1}^L|s_i\>_y$,
\be\label{eq:x_init}
\begin{split}
&\phi(0) = |\phi_y\> = (-\frac {is_1}2,0,\dots,\frac {is_1}2,0,\dots),\\
&\rho(0) = \rho_y(\vec s) \equiv 
\frac 14\left(\begin{matrix}
{\rm diag}_{3}(\vec d,2,\vec d)&{\rm diag}_{3}(\vec d,0,-\vec d)\\
{\rm diag}_{3}(-\vec d,0,\vec d)& {\rm diag}_{3}(-\vec d,2,-\vec d)
\end{matrix}\right),
\end{split}
\ee
Finally the expression for system initially polarized in z basis $|\Psi\> = \Motimes_{i=1}^L|s_i\>_z^{\otimes L}$, is
\be\label{eq:y_init}
\begin{split}
&\phi(0) = 0,\\
&\rho_z(\vec s,0) = 
\frac 12\left(\begin{matrix}
1+{\rm diag}(\vec s)&0\\
0& 1-{\rm diag}(\vec s)
\end{matrix}\right),
\end{split}
\ee
Using these initial values and the operator in Eq.\eqref{eqs:single_particle_u}, it is possible to compute the values of $\phi(t_n)$ and $\rho(t_n)$ at any given time $t_n$. Then, these values can be used to find the SR parameters.

For example, we derive the value of SR parameter for $\alpha$-polarization of the first spin given initially it is $\beta$-polarized, $\alpha, \beta = x,y$, as
\be\label{eq:Cab_calc}
\begin{split}
C_{\alpha\beta,1} = \Bigl|\sum_{i}\<\Gamma^\pi_i\>\Tr(\Gamma^\pi_i\sigma^\alpha_1)\Bigl| = \sum_{i=1,2}\<\phi_\alpha|\varphi^i_\pi\>\<\varphi^i_\pi|\phi_\beta\>
\end{split}
\ee
where $|\varphi^i_\pi\>$ are $\pi$ quasienergy eigensates of the single-fermion Floquet unitary in Eq. \eqref{eq:single_fermion_tranform}, and $|\phi_\alpha\>$ are defined in Eqs. \eqref{eq:x_init}-\eqref{eq:y_init}.

The expression for $z$-polarization is different. It can be found as 
\be\label{eq:Czz_calc}
\begin{split}
C_{zz,1} = \biggl|\sum_{i,j=1,2}\<1|\phi^i_0\>\<\phi^i_0|\rho_z(\vec s)|\phi^j_\pi\>\<\phi^j_\pi|1\>+{\rm h.c.}\biggl|
\end{split}
\ee
while the expression for $C_{zx,1}$ and $C_{zy,1}$ vanish. The expessions from Eqs.\eqref{eq:Cab_calc}-\eqref{eq:Czz_calc} are plotted in Fig. \ref{fig:full_picture1}c.

To obtain the expectation values for the rest of the qubits, one can use the Majorana basis $\gamma_i = c_i+c_i^\dag$, $\gamma_{2i+1} = -i(c_i-c_i^\dag)$ and the two-point correlation function
\be
K_{ij}(t) = i\<\Psi|\gamma_i(t)\gamma_j(t)|\Psi\>.
\ee
The evolution of the matrix $K$ can be connected to the evolution of SPDM by
\be
K(t) = R\;\rho(t) \;R^T, 
\qquad R = 
\left(\begin{matrix}
1&1\\
-i&i
\end{matrix}\right)\otimes I
\ee

The matrix $K$ can be used to connect single-particle excitations with spin observables.  Let $I$ and $J$ be two subsets of indices with increasing order, then one defines  $A_{IJ}$ as the matrix whose elements are $A_{ij}$ with $i\in I$, $j\in J$. Then the time-dependent expectation of $x$-polarizations writes
\be
\<\sigma^{x}_i\> = \<\Psi|\mathcal P_{z,i}(t) \gamma_{2i}(t)|\Psi\> = {\rm Pf}\, K_{II}, \qquad I = \{1,\dots, 2i\}
\ee
where $\mathcal P_i = \prod_{k=1}^{i-1}(-\sigma^z_k)$ is a string operator. 
Similarly, one calculates
\be
\<\sigma^{y}_i\> = \<\Psi|\mathcal P_{z,i}(t) \gamma_{2i+1}(t)|\Psi\> = {\rm Pf}\,K_{II},
\ee
where $I = \{1\dots,2i-1,2i+1\}$, and
\be
\<\sigma^z_i\> = i\<\Psi|\gamma_{2i}(t)\gamma_{2i+1}(t)|\Psi\>= K_{2i,2i+1}(t).
\ee
Using these time-dependent expression, we derive the dynamics shown in Fig. \ref{fig:spin_oscillations} panels a-c.

\begin{figure*}[t!]
    \centering\includegraphics[width=0.9\textwidth]{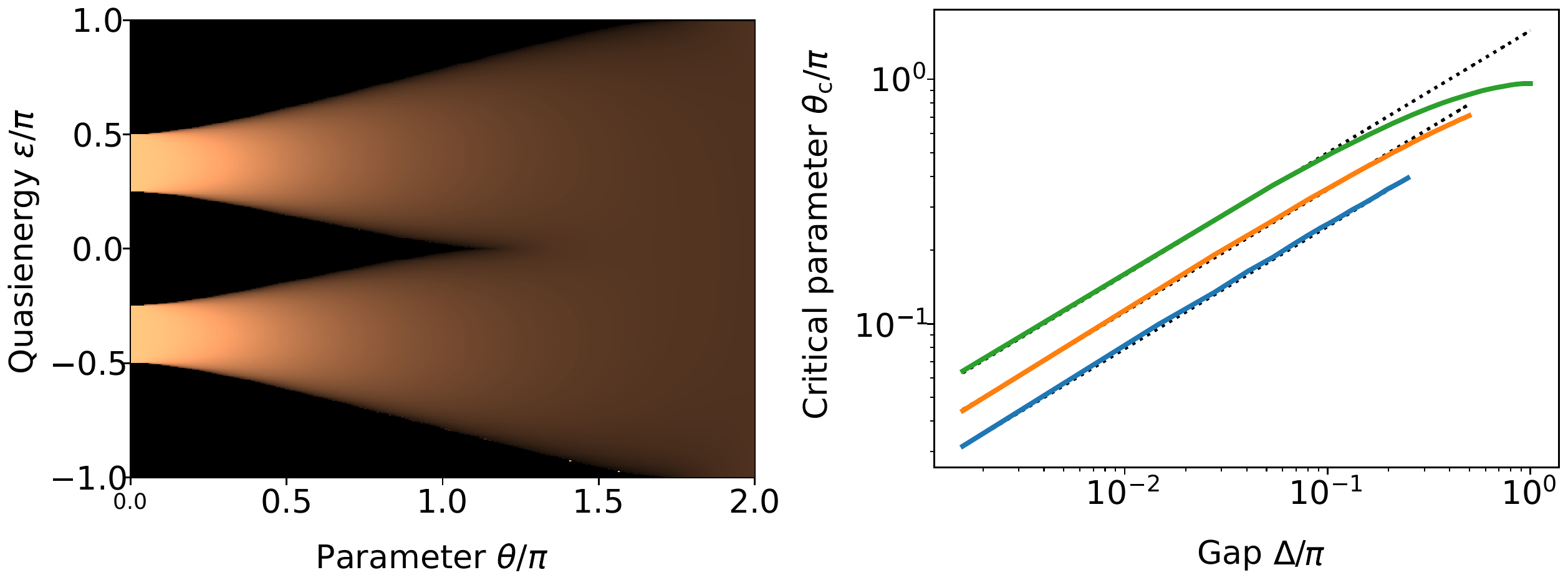}
\caption{\textbf{Floquet band structure.} 
\textbf{a.} The two-band spectral structure of the unitary operator $V'$ in Eq. \eqref{eq:quasiparticle_rot} given the density of states of $V$ and $U$ as in Eqs. \eqref{eq:V_dos}-\eqref{eq:U_dos}. The gap closes at critical $\theta_c$ which depends on $G$ and $\Delta$. \textbf{b.} The logarithmic scale plot of $\theta_c$ along with linear approximation (dashed lines). The linear approximation for most of curves is given by $\theta_c = c\sqrt{G\Delta}$ almost for all values of gap $\Delta$ and the bandwidth $T$, where $c = \sqrt{5/2}$ for the case we study here.}
\label{fig:fpt}
\end{figure*}

\subsection{Section 5: Stability of the gap (p.4)}

The Floquet Hamiltonian in presence of weak interactions preserves the modes $\tilde\psi_k = \sum_k \gamma_{kk'}\psi_{k'}+O(\lambda^2)$, where $U_{kk'}$ are parameters depending on $\lambda$ and the type of discrete disorder, if present. The modes operators $\tilde \psi_k$ in this linear approximation, in contrast to $\psi_k$ in Eq.\eqref{eq:sf_modes}, represent not real particles but quasiparticles with lifetime depending on the neglected $O(\lambda^2)$ part. Then the corresponding Floquet operator is characterized by a single fermion unitary matrix $V'$ (see Eq.$\,$\eqref{eq:single_fermion_tranform}), which obeys
\be \label{eq:quasiparticle_rot}
V' =  UV,
\ee
where $U$ is a interaction correction unitary operator and $V$ is a single fermion unitary corresponding to the non-interacting system.

The structure of the unitary $U$ is unknown, therefore we approximate its eigenvectors as Haar and restrict its eigenvalues (mod 2) to be such that $\|\log U\|\leq \theta$, where $\theta$ is a maximum mixing angle. This makes $U$ and $V$ free independent and we can use an imaginary time version of the S-transform in free probability theory. 

As a example, let us consider the normalized density of states as a function of quasienergy $\ve\in[-\pi,\pi]$ for the original non-interacting system unitary operator $V$ to be equal to
\be\label{eq:V_dos}
\rho_V(\ve) = 
\begin{cases}
(G-\Delta)^{-1}, \qquad \Delta/2\leq |\ve| \leq G/2\\
0, \qquad {\rm otherwise}
\end{cases}
\ee
This expression is simplification band structure for one shown in Fig. \ref{fig:full_picture1}c in the main text. Despite being not exact, it allows us understand qualitatively the effect of random unitary rotation in Eq. \eqref{eq:quasiparticle_rot}. 

Let us also assume that the density of states for the unitary $U$ is
\be\label{eq:U_dos}
\rho_U(\ve) = 
\begin{cases}
\theta^{-1}, \qquad &|\ve|<\theta/2 \\
0, \qquad &{\rm otherwise}
\end{cases}
\ee
 The parameter $\theta\to0$ represents the case $U = I$, while $\theta=2\pi$ corresponds to $U$ being a random unitary by Haar measure.

The density of states can be obtained from Herglotz transform:
\be\label{eq:Hertholtz}
h(z) = \int \frac{e^{i\ve}+z}{e^{i\ve}-z}\rho(\ve)d\ve
\ee
This can be inverted to obtain the density of states:
\be
\rho(\ve) = \frac 1{2\pi}\lim_{\xi\to+0}{\rm Re}\,h(e^{-i\ve-\xi}),
\ee
In particular, the Herglotz transform for the product in Eq.\eqref{eq:quasiparticle_rot} can be obtained by solving simultaneously the equations \cite{vasilchuk2001law}
\be
\label{eq:vasilchuk_method}
\begin{split}
&h^2(z) = 1+4z\Delta_1(z)\Delta_2(z)\\
&h(z) = h_1\left(\frac{2z\Delta_1(z)}{1+h(z)}\right)\\
&h(z) = h_2\left(\frac{2z\Delta_2(z)}{1+h(z)}\right)
\end{split}
\ee
for the class of function $h(z)$ which are analytic for $|z| < 1$ and satisfy:
\be\label{eq:restrictions}
\begin{split}
{\rm Re}\,h(z)>0,\qquad |h(z)-1|\leq \frac{2|z|}{1-|z|},
\end{split}
\ee
 while the functions must obey $\Delta_1(z)$, and $\Delta_2(z)$
 \be
 |\Delta_{1,2}(z)|\leq \frac{1}{1-|z|}
 \ee
 for $\text{for}\;|z|<1$.
 
The expression in Eq.\eqref{eq:vasilchuk_method} can be replaced by an implicit expression
\be\label{eq:main_equation_sol}
h(z) = h_V\Bigl(z\frac{h(z)-1}{h(z)+1}\frac 1{h_U^{-1}(h(z))}\Bigl)
\ee 
which should be solved for the function $h(z)$ for given $z$. The solution can be obtained numerically by Newton's method or simply by iterations.

The Hertglotz transformations $h_V(z)$ and $h_U(z)$ can be calculated analytically, which yields
\be\label{eq:hertholtz_V}
h_V(z) = -1+\frac{2i}{G-\Delta}\log\biggl(\frac{z^2-az+b}{z^2-a^*z+b^*}\biggl),
\ee
and
\be
h_U(z) = -1+\frac{2i}\theta\log\biggl(\frac{e^{-i\theta/2}-z}{e^{i\theta/2}-z}\biggl),
\ee
where $a = e^{-i(G-\Delta)/2}+e^{i\Delta/2}$, $b = e^{-i(G-\Delta)/2}$.

Functional inversion of the transform $h_U(z)$ can be written in the compact form
\be\label{eq:inv_hertholtz_U}
h_U^{-1}(w) = \frac{\sin[(w-1)\theta/4]}{\sin[(w+1)\theta/4]}
\ee
Combining Eqs. \eqref{eq:hertholtz_V} and \eqref{eq:inv_hertholtz_U} with Eq. \eqref{eq:main_equation_sol}, we obtain the numerical solution for the density of states as well as the gap. As seen from Fig. \ref{fig:fpt}, for given badnwidth $G$ and $\Delta$ the bandgap closes at a particular maximum mixing angle scaling as $\theta_c = c\sqrt{G\Delta}$, for some $c$. Assuming that for a small coupling term $\lambda$ the phase in Eq.\eqref{eq:U_dos} corresponds to $\theta\sim \lambda$, we conclude that the critical disorder closing the gap is $\lambda_c \sim \sqrt{GT}$.

\end{document}